      [1999/12/01 v1.4c Il Nuovo Cimento]
\documentclass{cimento}

\title{Spinning particles in General Relativity}
\author{F.~Cianfrani\from{ins:x}
                  \atque    
G.~Montani\from{ins:x}\from{ins:y}}
\instlist{\inst{ins:x} ICRA-International Center for Relativistic Astrophysics\\
Dipartimento di Fisica (G9), Universit\`a  di Roma, ``La Sapienza",\\
Piazzale Aldo Moro 5, 00185 Rome, Italy.
  \inst{ins:y} ENEA C.R. Frascati (Dipartimento F.T.P.),\\ via Enrico Fermi 45, 00044 Frascati, Rome, Italy.}
\PACSes{{04.90.+e}\hspace{0.3cm}{04.50.+h}}
\begin{document}

\maketitle

\begin{abstract}
We analyze the behavior of a spinning particle in gravity, both from a quantum and a classical point of view. We infer that, since the interaction between the space-time curvature and a spinning test particle is expected, then the main features of such an interaction can get light on which degrees of freedom have physical meaning in a quantum gravity theory with fermions. Finally, the dimensional reduction of Papapetrou equations is performed in a 5-dimensional Kaluza-Klein background and Dixon-Souriau results for the motion of a charged spinning body are obtained.
\end{abstract}

\section{Quantum features of the interaction between spin and gravity}

The interaction between gravity and fundamental particles is still an open issue of our knowledge. Even if the final answer to this problem must be given by a quantum formulation for the gravitational field, nevertheless we expect that a good effective description (for energy scales much lower that Planck's one) could come out from the study of particles dynamics living on a curved space-time. But also this is a highly non-trivial task, because, in a classical picture a free moving particle follows geodesics lines, but there are no unambiguous indications for the motion of spinning particles; in fact, in this last case, the classical dynamics must be inferred as classical limit of a relativistic quantum mechanical equation.\\ 
In this respect, let us consider the Dirac equation on a curved space-time
\begin{equation} 
\gamma^\mu D_\mu\psi=0\qquad D_\mu\psi=\bigg(\partial_\mu-\frac{i}{2}\omega^{ab}_\mu\Sigma_{ab}\bigg)\psi 
\end{equation}
being $\omega^{ab}_\mu$ spin connections and $\gamma^\mu$ Dirac matrices on the given space-time.
If we square the Dirac operator, the following second order equation arises
\begin{equation}
g^{\mu\nu}D_{\mu}D_{\nu}\psi-\frac{1}{4}R\psi=0
\end{equation}
being $R$ the scalar curvature associated to the manifold.
Therefore, we expect a non-trivial spin-curvature coupling, since the last equation differs significantly from the Klein-Gordon one (with a non-minimal curvature coupling), describing the dynamics of a spinless particle
\begin{equation}
g^{\mu\nu}\partial_{\mu}\partial_{\nu}\phi-\frac{1}{4}\xi R\phi=0
\end{equation}
because of the presence of spinor connections into covariant derivatives $D_{\mu}$.\\
Furthermore, the back-reaction of spinors on the space-time outlines that they have a direct effect on the geometry, in terms of the appearance of a non-vanishing torsion: by varying the Einstein-Dirac action (written in a Palatini-like formulation) with respect to $\omega^{ab}_{\mu}$, the following second Cartan structure equation is obtained
\begin{equation}
\partial_{[\mu}e^a_{\nu]}+\omega^a_{[\mu |b|}e^b_{\nu]}=\frac{1}{4}\epsilon^{a}_{\phantom1 bcd}e^b_\nu e^c_\mu J_A^d\qquad J_A^a=\bar{\psi}\gamma^a\gamma_5\psi,
\end{equation}
admitting the following solution $\omega^{ab}_\mu={}^0\omega^{ab}_\mu+\frac{1}{4}\epsilon^{ab}_{\phantom1\phantom2 cd}e^c_\mu J_A^d$, with ${}^0\omega^{ab}_\mu=e^b_\nu\nabla_\mu e^{a\nu}$.

Therefore, we can rewrite the Einstein-Dirac action as follows
\begin{equation}
S=-\frac{1}{2}\int\bigg(R+i\bar{\psi}\gamma^\mu{}^{(0)}D_\mu\psi+c.c.-\frac{3}{8}\eta_{ab}J^a_AJ^b_A\bigg)ed^4x;
\end{equation}
we emphasize the appearance in this reduced action of an interaction term containing four fermions, which, in general, provides non renormalizable contributions to quantum amplitudes.\\
Thus, the analysis above singlets out a peculiar property of spinors back-reaction on the space-time; in fact, differently from ordinary matter, spinors induce not only a small change to the metric field, but also a non-vanishing torsion field; therefore, the classical limit of spinors dynamics is expected to be no longer a geodesics line, as far as a spin notion is recovered. Of course, the spin-curvature coupling can survive in a classical limit only if the metric field changes significantly on the Compton scale of fermions; in this case, an effective classical theory describing the non geodesics motion of a spinning particle can have a predictive character. However, we emphasize that, taking such kind of classical limit contains a good degree of ambiguity and a unique classical effective theory does not arise. 

\section{Classical formulation for spinning particles}

The result above presented, about the relation between spin and gravity, has its classical counterpart in the effect which the presence of an angular momentum has on the dynamics of a macroscopic body. In particular, the well-known Papapetrou work \cite{pp} was devoted to study the motion of a body in General Relativity on a fixed background.\\
In particular, in his analysis Papapetrou performed a multipoles expansion such that, if the first two moments are the only non-vanishing ones, then a characterization of the internal structure of the body takes place in terms of what he called the spin tensor (indeed this notion is recovered as a point-like limit of a classical angular momentum), i.e.
\begin{equation}
S^{\mu\nu}=\int_{\tau}\delta x^{\mu}T^{\nu 0}-\int_{\tau}\delta x^{\nu}T^{\mu 0}.
\end{equation} 
with $\tau$ denoting the hypersurface at equal time, $T^{\mu\nu}$ the energy momentum tensor of the body and $\delta x^\mu$ the displacement with respect to the center of mass.\\
So Papapetrou recognized the pole-dipole approximation as able to describe a spinning body coupled to gravity and he found the following system of dynamical equations
\begin{equation}
\left\{\begin{array}{c}\frac{D}{Ds}P^{\mu}=\frac{1}{2}R_{\nu\rho\sigma}^{\phantom1\phantom2\phantom3 \mu} S^{\nu\rho}u^{\sigma}\quad\\\\
\frac{D}{Ds}S^{\mu\nu}=P^{\mu}u^{\nu}-P^{\nu}u^{\mu}\\\\
P^{\mu}=mu^{\mu}-\frac{DS^{\mu\nu}}{Ds}u_{\nu}\label{pe}\quad\end{array}\right.
\end{equation}
being $u^\mu$ the 4-velocity, $m$ the rest mass of the body and $R_{\nu\rho\sigma}^{\phantom1\phantom2\phantom3 \mu}$ the Riemann tensor.\\
Thus, a spinning body has a recognizable interaction with the gravitational curvature, through its spin tensor, and it does not follow a geodesics motion; in this scheme, the particle retains a test character, but it acquires an internal structure, due to its spin, which prevents it a free motion even in absence of ``external'' fields.\\ 
To closed the system (\ref{pe}), we must add one of the following additional conditions
\begin{equation}
S^{\mu 0}=0\qquad S^{\mu\nu}u_{\nu}=0\qquad S^{\mu\nu}P_{\nu}=0
\end{equation}
known in literature as the Papapetrou, Pirani and Tulczyjew conditions, respectively.\\
In the weak field approximation, these conditions provide equivalent results and therefore current experiments cannot select among them.\\
In this scenario, the introduction of an external electro-magnetic field $F_{\mu\nu}$ is due to Dixon and Souriau \cite{dix,sou}, through the following extension of the Papapetrou paradigm
\begin{equation}
\left\{\begin{array}{c}\frac{D}{Ds}P^{\mu}=\frac{1}{2}R_{\nu\rho\sigma}^{\phantom1\phantom2\phantom3 \mu} S^{\nu\rho}u^{\sigma}+qF^{\mu}_{\phantom1\rho}u^{\rho}+\frac{1}{2}M^{\sigma\nu}\nabla^{\mu}F_{\sigma\nu}\\\\
\frac{D}{Ds}S^{\mu\nu}=P^{\mu}u^{\nu}-P^{\nu}u^{\mu}-M^{\mu\rho}F_{\rho}^{\phantom1\nu}-M^{\nu\rho}F_{\rho}^{\phantom1\mu}\\\\
\frac{dq}{ds}=0\qquad\qquad\qquad\qquad\qquad\qquad\end{array}\right.\label{s3}
\end{equation}
$q$ and $M^{\mu\nu}$ being the electric charge and the electro-magnetic moment, respectively.\\
Our aim is to show that, for a particle of charge $q$ and mass $m$, the Dixon-Souriau equations can be obtained as dimensional reduction of the system (\ref{pe}) in a 5-dimensional Kaluza-Klein background, so emphasizing that the geometrization of the electro-magnetic interaction preserves particles dynamics, up to the dipole order.\\
The space-time of the Kaluza-Klein \cite{2,3} theory is the direct sum of a 4-dimensional one $V^4$ and of an extra-dimension compactified to a circle of Planck length scale, such that it is un-observable. The corresponding metric field takes the form
\begin{equation}
\label{c1}
j_{AB}=\left(\begin{array}{cc}g_{\mu\nu}-e^{2}k^{2}A_{\mu}A_{\nu}
& -ekA_{\mu} \\
-ekA_{\nu}
& -1\end{array}\right)
\end{equation} 
$g_{\mu\nu}=g_{\mu\nu}(x^{\rho})$ being the 4-dimensional metric, while $A_\mu=A_\mu(x^{\rho})$ can be interpret as the electro-magnetic vector potential. In fact, if we split the Einstein-Hilbert action in such a space-time
\begin{equation}
\label{k1}S=-\frac{c^{4}}{16\pi {}^{(5)}\!G}\int_{V^{4}\otimes S^{1}}\sqrt{-j}{}^{(5)}\!R d^{4}xdx^{5}
\end{equation} 
the Maxwell Lagrangian density for $A_\mu$ comes out 
\begin{equation}
S=-\frac{c^{3}}{16\pi}\frac{2\pi R}{{}^{(5)}\!G}\int_{V^{4}}
\sqrt{-g}\bigg[R+\frac{e^{2}k^{2}}{4}F_{\mu\nu}F_{\rho\sigma}g^{\mu\rho}g^{\nu\sigma}\bigg]d^{4}x.
\end{equation}  
In this sense, we recognize electro-magnetic degrees of freedom as components of the 5-dimensional metric.\\
Within such a space-time picture, one can show \cite{6,7} that the dimensional reduction of a geodesics motion gives the right trajectory followed by charge test-particles. 
Let us now consider Papapetrou equations in a KK space-time, i.e. in (\ref{pe}) we replace indices $\mu,\nu$ by $\Omega,\Pi=0,1,2,3,5$ and the spin tensor $S^{\mu\nu}$ by the corresponding 5-dimensional one $\Sigma^{\Omega\Pi}$.
Once the following identifications stand 
\begin{eqnarray}
\Sigma^{\mu\nu}=S^{\mu\nu}\qquad S_{\mu}=\Sigma_{5\mu}\qquad q=2m\sqrt{G}u_5\\
M^{\mu\nu}=\frac{1}{2}ek(S^{\mu\nu}u_{5}+u^{\mu}S^{\nu}-u^{\nu}S^{\mu})
\end{eqnarray}
we recover Dixon-Souriau equations \cite{6,7},
with the extra-components of the spin tensor ($\Sigma_{5\mu}$) describing a non-vanishing electric dipole moment.\\

\section{Ashtekar formalism and chirality eigenstates}

The relation between spinning particles and gravity can have very interesting consequences in our understanding of the physical degrees of freedom concerning a quantum gravity theory in presence of fermions.\\
Let us consider the framework of Ashtekar formulation for the gravitational field dynamics \cite{rov} \cite{ash}, which is the most promising tool for a consistent Quantum Gravity theory; it is based on rewriting the Einstein-Hilbert action in terms of Ashtekar connections $A^k_\mu$ and $A^{*k}_\mu$, i.e.
\begin{equation} 
A^k_\mu=\omega^{0k}_\mu-\frac{i}{2}\epsilon^k_{ij}\omega^{ij}_\mu\qquad A^{*k}_\mu=\omega^{0k}_\mu+\frac{i}{2}\epsilon^k_{ij}\omega^{ij}_\mu
\end{equation}
which are the self-dual and antiself-dual parts of the Lorentz connection, respectively. Hence, the gravitational action can be split in two independent terms, one written in $A^k_\mu$ and one in $A^{*k}_\mu$, as follows
\begin{eqnarray}
S=\frac{1}{2}\int e^\mu_a e^\nu_b R^{ab}_{\mu\nu} e d^4x=\int e^\mu_a e^\nu_b (F^k_{\mu\nu}+F^{*k}_{\mu\nu})e d^4x
\end{eqnarray}
being $F^k_{\mu\nu}$ and $F^{*k}_{\mu\nu}$ the curvature of Ashtekar connections
\begin{eqnarray}
F^k_{\mu\nu}=\partial_\mu A^k_\nu-\partial_\nu A^k_\mu-\frac{i}{2}\epsilon^k_{ij}A^i_\mu A^j_\nu\quad F^{*k}_{\mu\nu}=\partial_\mu A^{*k}_\nu-\partial_\nu A^{*k}_\mu+\frac{i}{2}\epsilon^k_{ij}A^{*i}_\mu A^{*j}_\nu. 
\end{eqnarray}
The coupling with spinors suggests a physical interpretation of such connections; by well-known properties of Dirac matricies, we have
\begin{eqnarray*}
\frac{i}{2}\bigg[(\bar{\psi}\gamma^\mu\Sigma_{ab}\psi+\bar{\psi}\Sigma_{ab}\gamma^\mu\psi)\omega^{ab}_\mu\bigg]=\frac{i}{2}\bigg[(\bar{\psi}\gamma^\mu \sigma_k\psi_R+\bar{\psi}_L\sigma_k\gamma^\mu \psi) A^k_\mu-(\bar{\psi}\gamma^\mu \sigma_k\psi_L+\bar{\psi}_R\sigma_k\gamma^\mu \psi) A^{*k}_\mu\bigg]
\end{eqnarray*}
therefore $A^k_\mu$ turns out to be the connection associated to right-handed particles and antiparticles, and $A^{*k}_\mu$ has analogous role for left-handed ones. Hence, until $P$ and $CP$ invariance holds, we have a full correspondence, through these symmetries, between currents associated to $A^k_\mu$ and $A^{*k}_\mu$; but, as soon as $P$ and $CP$ are simultaneously violated, these currents would acquire an independent physical nature. As a relevant consequence, in this context basic connection variables must be identified in $A^k_\mu$ and $A^{*k}_\mu$, respectively. From a physical point of view, this would imply that the $SU(2)$ group becomes a more fundamental symmetry with respect to the Lorentz one. But here the question about why experiments provide evidence for the Lorentz symmetry, instead of a fundamental $SU(2)$ one, would naturally arise. In order to precise the physical characterization of such fundamental role played by the $SU(2)$ group, we propose the following symmetry breaking scenario. Before particles acquire masses, $\psi_L$ and $\psi_R$ are independent and there is no objection in taking $A^k_\mu$, $A^{*k}_\mu$ as different physical fields, if $CP$ is violated ($P$ violations being provided by the standard Electro-Weak model chirality). But, when the energy drops down to scales at which the spontaneous symmetry breaking process occurs, then we get massive particles, which means that the dynamics of $\psi_L$ and $\psi_R$ is correlated, so a new Lorentz connection $\omega^{ab}_\mu$ arises. Of course, the present observation of the leptons as individual particles requires, in this scheme, a fundamental interaction, which confines their left and right components. Therefore, just at sufficiently high energies we can deal with independent Ashtekar connections and independent left and right fermions.


\begin{thebibliography}{99}

\bibitem{pp}
\BY{Papapetrou~A.}\IN{Proc. Roy. Soc. London}{A209}{1951}{248}

\bibitem{dix}
\BY{Dixon~W.~G.}\IN{Il Nuovo Cimento}{A XXXIV, $n^{o}$ 2}{1964}{317}

\bibitem{sou} 
\BY{Souriau~J.~M.}\IN{Ann. Inst. H. Poincaré}{A XX, $n^{o}$ 4}{1974}{22}

\bibitem{2}
\BY{Kaluza~T.}\IN{Sitzungseber. Press. Akad. Wiss. Phys. Math. Klasse}{K1}{1921}{966}

\bibitem{3}
\BY{Klein~O.}\IN{Nature}{118}{1926}{516}

\bibitem{6}
\BY{Cianfrani~F. \atque Marrocco~A. \atque Montani~G.}
\IN{Int. J. Mod. Phys}{D14, 7}{2005}{1195}

\bibitem{7}
\BY{Cianfrani~F. \atque Milillo~I. \atque Montani~G.} \TITLE{Dixon-Souriau equations from a 5-dimensional spinning particle in a Kaluza-Klein framework}, submitted to Phys. Lett. A

\bibitem{rov}
\BY{Ashtekar~A.} \IN{Phys. Rev. D}{36}{1987}{1587}

\bibitem{ash}
\BY{Ashtekar~A.} \IN{Phys. Rev. Lett.}{57}{1986}{2244}


\end{thebibliography}
\end{document}